# A huge explosion in the early Universe

G. Cusumano[1], V. Mangano[1], G. Chincarini[2,3], A. Panaitescu[4], D.N. Burrows[5], V. La Parola[1], T. Sakamoto[6,7], S. Campana[2], T. Mineo[1], G. Tagliaferri[2], L. Angelini[6], S.D. Barthelemy[6], A.P. Beardmore[8], P.T. Boyd[6], L.R. Cominsky[9], C. Gronwall[5], E.E. Fenimore[4], N. Gehrels[6], P. Giommi[10], M. Goad[8], K. Hurley[11], J.A. Kennea[5], K.O. Mason[12], F. Marshall[6], P. Mèszàros[5,13], J.A. Nousek[5], J.P. Osborne[8], D.M. Palmer[4], P.W.A. Roming[5], A. Wells[8], N.E. White[6], B. Zhang[14]

[1]*INAF-Istituto di Astrofisica Spaziale e Fisica Cosmica di Palermo, Via Ugo La Malfa 153, 90146 Palermo, Italy,*

[2]*INAF -- Osservatorio Astronomico di Brera, Via Bianchi 46, 23807 Merate Italy*

[3]*Università degli studi di Milano-Bicocca, Dip. di Fisica, Piazza delle Scienze 3, I-20126 Milan, Italy,*

[4]*Los Alamos National Laboratory, P.O. Box 1663, Los Alamos, NM 87545, USA*

[5]*Department of Astronomy & Astrophysics, Pennsylvania State University, University Park, PA 16802 USA.*

[6]*NASA/Goddard Space Flight Center, Greenbelt, MD 20771, USA*

[7]*National Research Council, 2001 Constitution Avenue, NW, TJ2114, Washington, DC 20418, USA*

[8]*Department of Physics and Astronomy, University of Leicester, University Road, Leicester LE1 7RH, UK*

[9]*Department of Physics and Astronomy, Sonoma State University, Rohnert Park, CA 94928-3609, USA*

[10]*ASI Science Data Center, via Galileo Galilei, 00044 Frascati, Italy*

[11]*UC Berkeley Space Sciences Laboratory, Berkeley, CA 94720-7450*

[12]*MSSL, University College London, Holmbury St. Mary, Dorking, RH5 6NT Surrey, UK*

[13]*Department of Physics, Pennsylvania State University, PA 16802, USA*

[14]*Department of Physics, University of Nevada, Box 454002, Las Vegas, NV 89154-4002, USA*

Long gamma-ray bursts (GRBs) are bright flashes of high-energy photons that can last up to tens of minutes; they are generally associated with galaxies that have a high rate of star formation and probably arise from the collapsing core of massive stars which produce highly relativistic jets (collapsar model[1]). Here we describe γ- and X-ray observations of the most distant GRb ever observed (GRB050904): its redshift[2,3] of 6.29 means that this explosion happened even as far back as 12.8 billion years ago, corresponding to a time when the Universe was just 890 million years old, close to the reionization era[4]: this means that not only have stars formed within this short period of time since the Big Bang, but also enough time has elapsed for them to have evolved and collapsed into black holes.

GRB050904 triggered the Burst Alert Telescope (BAT) onboard the Swift[5] satellite on 2005 September 4 at 01:51:44 UT and the spacecraft quickly slewed to it to allow observations by the X-ray Telescope (XRT)[6,7], which observed it up to 10 days after the burst onset. Figure 1 (top panel) shows the time history of the burst. Hereafter the GRB phenomenology is presented and discussed from the point of view of the source rest frame

The BAT light curve shows three main peaks: two ~ 2 second long peaks at T+3.8 and at T+7.7 seconds, a main long-lasting peak at ~ T+13.7 seconds and weak peak at ~ T+64 seconds, where T is the time of the burst onset. The early XRT light curve shows a steep power law decay with an index of -2.07 ± 0.03 with two flares superimposed at T+64 seconds, coincident with the last peak of the BAT light curve, and T+170 seconds. Although interrupted by low Earth orbit observing constraints, the X-ray light curve reveals highly irregular intensity variations likely due to the presence of flares up to T+1.5 hours. At later times the flaring activity is not detected leaving only a residual emission $10^6$ times lower than the initial intensity.

The flares in the XRT light curve can be interpreted as late internal shocks related to central engine activity. In this scenario they would have the same origin as the prompt gamma-ray emission[8,9,10], requiring that the central engine remains active for at least 5000 seconds, consistent with the collapsar model[1].

Spectral analysis was performed by selecting time intervals corresponding to characteristic phases of the light curve evolution. All spectra were well modelled by a single power law, with both Galactic and intrinsic absorption components in the case of the XRT spectra. Figure 1 (bottom panel) shows the evolution with time of the photon index Γ. The BAT spectra have Γ=-1.2. If we exclude the spectrum of the first XRT flare at T+64 seconds, the XRT photon indices show a clear decreasing trend from about -1.2 to about -1.8 in the first T+200 seconds. No further spectral evolution is present in later XRT data.

The overall phenomenology of GRB050904 is not peculiar with respect to other GRBs at lower redshift. This suggests that GRB explosions in the early and present Universe undergo similar physical mechanisms.

Based on the likely existence of Pop I/II stars in galaxies that were already metal-enriched at these high redshifts[11] we expect ~ 10% of all bursts detected by Swift to be located at $z \geq 5$. A higher percentage would require an additional contribution to the high redshift GRB population by metal-free Pop III stars, which have been shown to be viable GRB progenitors for long duration GRBs[11]. A more systematic search for GRB optical counterparts will increase the sample of these high redshift GRBs, allowing us to probe the existence of metal-free Population III massive stars.

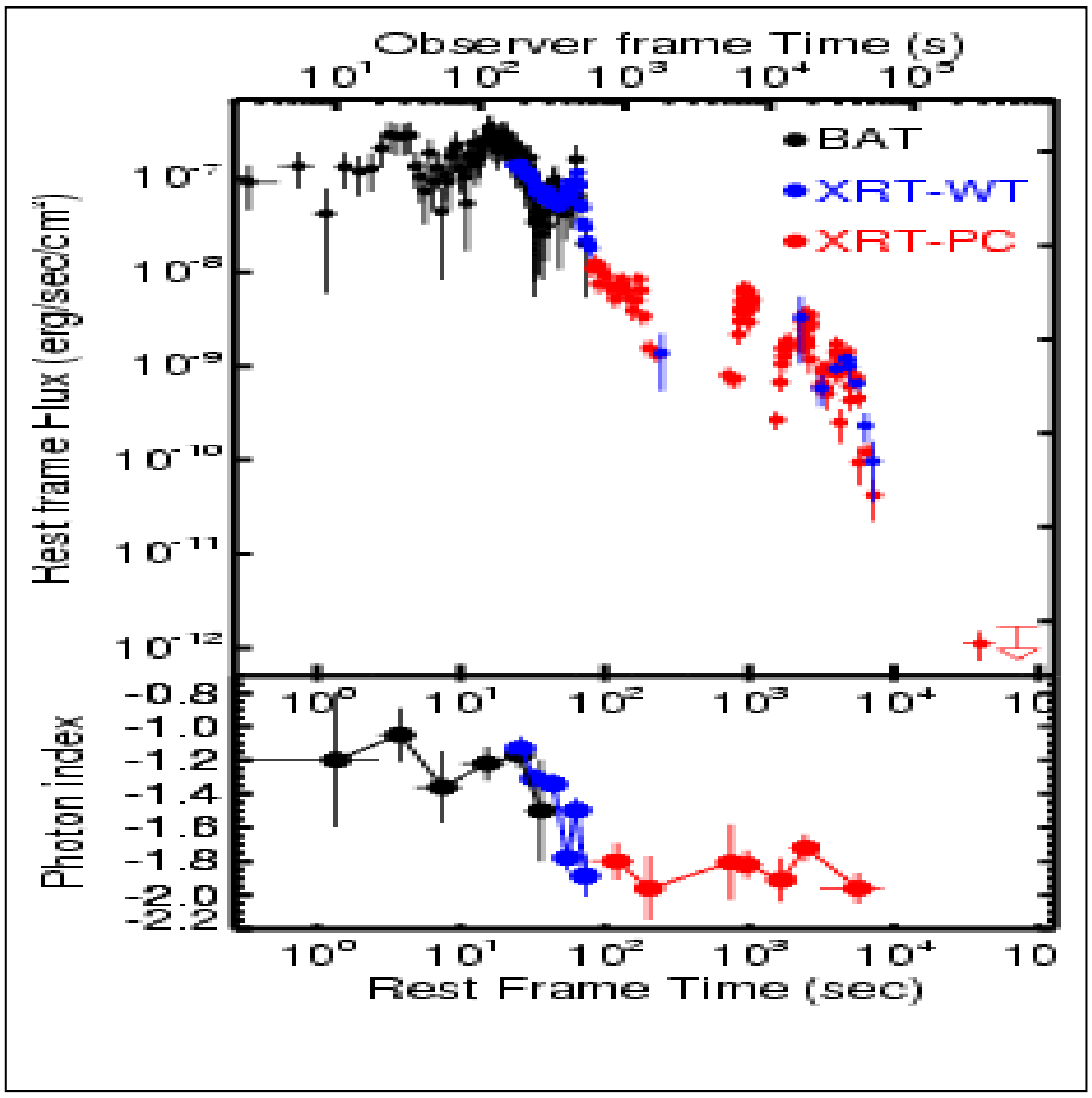

Figure 1. **Light curve and spectral evolution of GRB050904 as observed by the BAT and XRT**. WT is windowed timing mode data, and PC is photon counting data. The top panel shows the evolution of the GRB K-corrected 0.2-10 keV luminosity. The error bars are at 90% confidence level. The lower horizontal axis shows the time in seconds starting from the BAT trigger in the rest frame, obtained by applying the correction factor $(1+z)^{-1}$ to the observer frame time (upper horizontal axis). The gaps in the XRT-PC data correspond to the part of the orbit when the satellite was not observing this GRB. The bottom panel illustrates how the photon index $\Gamma$ of GRB050904 changes during the observation. The photon index is defined by the power law $F(E)=E^{\Gamma+1}$. The spectra were modelled using a power law with two absorbing components (Galactic and intrinsic).